\documentclass[twocolumn]{aastex631}

\newcommand{\lya}{Ly$\alpha$}
\newcommand{\oiii}{[O\,{\footnotesize III}]}

\newcommand{\kms}{km s$^{-1}$}
\newcommand{\myr}{$M_{\odot}$ yr$^{-1}$}
\newcommand{\msun}{$M_{\odot}$}
\newcommand{\mstar}{$M_{\star}$}

\newcommand{\kkmspc}{K~km~s$^{-1}$~pc$^2$}

\newcommand{\jykms}{Jy~km~s$^{-1}$}

\newcommand{\hi}{H\,{\footnotesize I}}
\newcommand{\hii}{H\,{\footnotesize II}}

\newcommand{\cii}{[C\,{\footnotesize II}]}

\newcommand{\lcii}{$L_{\rm [CII]}$}

\newcommand{\dlya}{$\Delta v_{\mathrm{Ly\alpha}}$}
\newcommand{\dxya}{$\Delta x_{\mathrm{Ly\alpha}}$}
\newcommand{\ewlya}{EW$_0(\rm Ly\alpha)$}
\newcommand{\muv}{$M_{\rm UV}$}
\newcommand{\mhi}{$M_{\rm{H\,\scriptscriptstyle{I}}}$}
\newcommand{\nhi}{$N_{\rm{H\,\scriptscriptstyle{I}}}$}

\newcommand{\gal}{Y002}

\shorttitle{A strong \lya\ and \cii\ emitter in the epoch of reionization}
\shortauthors{Valentino et al.}
\graphicspath{{./}{figures/}}

\begin{document}

\title{The archival discovery of a strong Lyman-$\alpha$ and \cii\ emitter at $z=7.677$}

\correspondingauthor{Francesco Valentino}
\email{francesco.valentino@nbi.ku.dk}

\author[0000-0001-6477-4011]{Francesco Valentino}
\affiliation{Cosmic Dawn Center (DAWN), Denmark}
\affiliation{Niels Bohr Institute, University of Copenhagen, Jagtvej 128, DK-2200 Copenhagen N, Denmark}

\author[0000-0003-2680-005X]{Gabriel Brammer}
\affiliation{Cosmic Dawn Center (DAWN), Denmark}
\affiliation{Niels Bohr Institute, University of Copenhagen, Jagtvej 128, DK-2200 Copenhagen N, Denmark}

\author[0000-0001-7201-5066]{Seiji Fujimoto}
\affiliation{Cosmic Dawn Center (DAWN), Denmark}
\affiliation{Niels Bohr Institute, University of Copenhagen, Jagtvej 128, DK-2200 Copenhagen N, Denmark}

\author[0000-0002-9389-7413]{Kasper E. Heintz}
\affiliation{Cosmic Dawn Center (DAWN), Denmark}
\affiliation{Niels Bohr Institute, University of Copenhagen, Jagtvej 128, DK-2200 Copenhagen N, Denmark}
\affiliation{Centre for Astrophysics and Cosmology, Science Institute, University of Iceland, Dunhagi 5, 107 Reykjav\'ik, Iceland}

\author[0000-0003-1614-196X]{John R. Weaver}
\affiliation{Cosmic Dawn Center (DAWN), Denmark}
\affiliation{Niels Bohr Institute, University of Copenhagen, Jagtvej 128, DK-2200 Copenhagen N, Denmark}

\author[0000-0002-6338-7295]{Victoria Strait}
\affiliation{Cosmic Dawn Center (DAWN), Denmark}
\affiliation{Niels Bohr Institute, University of Copenhagen, Jagtvej 128, DK-2200 Copenhagen N, Denmark}

\author[0000-0003-4196-5960]{Katriona M. L. Gould}
\affiliation{Cosmic Dawn Center (DAWN), Denmark}
\affiliation{Niels Bohr Institute, University of Copenhagen, Jagtvej 128, DK-2200 Copenhagen N, Denmark}

\author[0000-0002-3407-1785]{Charlotte Mason}
\affiliation{Cosmic Dawn Center (DAWN), Denmark}
\affiliation{Niels Bohr Institute, University of Copenhagen, Jagtvej 128, DK-2200 Copenhagen N, Denmark}

\author[0000-0002-4465-8264]{Darach Watson}
\affiliation{Cosmic Dawn Center (DAWN), Denmark}
\affiliation{Niels Bohr Institute, University of Copenhagen, Jagtvej 128, DK-2200 Copenhagen N, Denmark}

\author[0000-0003-4207-0245]{Peter Laursen}
\affiliation{Cosmic Dawn Center (DAWN), Denmark}
\affiliation{Niels Bohr Institute, University of Copenhagen, Jagtvej 128, DK-2200 Copenhagen N, Denmark}

\author[0000-0003-3631-7176]{Sune Toft}
\affiliation{Cosmic Dawn Center (DAWN), Denmark}
\affiliation{Niels Bohr Institute, University of Copenhagen, Jagtvej 128, DK-2200 Copenhagen N, Denmark}

\begin{abstract}
We report the archival discovery of Lyman-$\alpha$ emission from the bright ultraviolet galaxy \gal\ at $z=7.677$, spectroscopically confirmed by its ionized carbon \cii$\,$158$\,\mu$m emission line. The \lya\ line is spatially associated with the rest-frame UV stellar emission ($M_{\rm UV}\sim-22$, $2\times$ brighter than $M^{\star}_{\rm UV}$) and it appears offset from the peak of the extended \cii\ emission at the current $\sim1\arcsec$ spatial resolution. We derive an estimate of the unobscured of $\mathrm{SFR_{UV}}=(22\pm1)$ \myr\ and set an upper limit of $\mathrm{SFR_{IR}}<15$ \myr\ from the far-infrared wavelength range, which globally places \gal\ on the SFR(UV+IR)-\lcii\ correlation observed at lower redshifts. 
In terms of velocity, the peak of the \lya\ emission is redshifted by $\Delta v_{\mathrm{Ly\alpha}}\sim500$ \kms\ from the systemic redshift set by \cii\ and a high-velocity tail extends to up to $\sim1000$ \kms. The velocity offset is up to $\sim3.5\times$ higher than the average estimate for similarly UV-bright emitters at $z\sim6\text{--}7$, which might suggest that we are witnessing the merging of two clumps. A combination of strong outflows and the possible presence of an extended ionized bubble surrounding \gal\ would likely facilitate the escape of copious \lya\ light, as indicated by the large equivalent width \ewlya$=24^{+5}_{-6}$ \AA. Assuming that \cii\ traces the neutral hydrogen, we estimate a \hi\ gas fraction of \mhi/\mstar$\gtrsim8$ for \gal\ as a system and speculate that patches of high \hi\ column densities could contribute to explain the observed spatial offsets between \lya\ and \cii\ emitting regions. The low dust content, implied by the non-detection of the far-infrared continuum emission at rest-frame $\sim160$ $\mu$m, would be sufficient to absorb any potential \lya\ photons produced within the \cii\ clump as a result of large \hi\ column densities.
\end{abstract}



\section{Introduction}
\label{sec:introduction}
The direct confirmation and the physical characterization of a large number of galaxies at $z>6\text{--}7$ is an important step towards a complete understanding of the process of reionization of the intergalactic medium (IGM) at this epoch, its main drivers, timeline, and topology \citep{loeb_2001, dayal_2018}.
Thus far, the workhorses of the spectroscopic campaigns targeting high-redshift sources have been the Lyman-$\alpha$ emission (\lya\ at $1215.16$ \AA) and, more recently, the bright far-infrared (FIR) cooling lines of ionized carbon (\cii\ at $157.74$ $\mu$m) and oxygen (\oiii\ at $51.82$ and $88.36$ $\mu$m). The combination of these lines from warm (\lya, \oiii) and cold (\cii) phases of the interstellar medium (ISM), together with the modeling of the stellar and dust-emitted light, also gives us the opportunity to start deciphering the earliest phases of galaxy growth. The resonant \lya\ emission is a primary tool of investigation, being sensitive to the onset of star formation, its feedback on the surrounding ISM and IGM, and to what extent their main component, hydrogen, is neutral or ionized. The addition of non-resonant FIR transitions not only allows us to pinpoint the systemic redshifts inaccessible with \lya, but offers a complementary insight into the metallicity, ionization, and dynamics of the ISM without being hampered by dust absorption as strongly as rest-frame optical lines. Moreover, their underlying FIR continuum emission tracks the first dust grains formed in the Universe.\\ 
\begin{figure}
    \vspace*{0.3cm}
    \centering
    \includegraphics[width=\columnwidth]{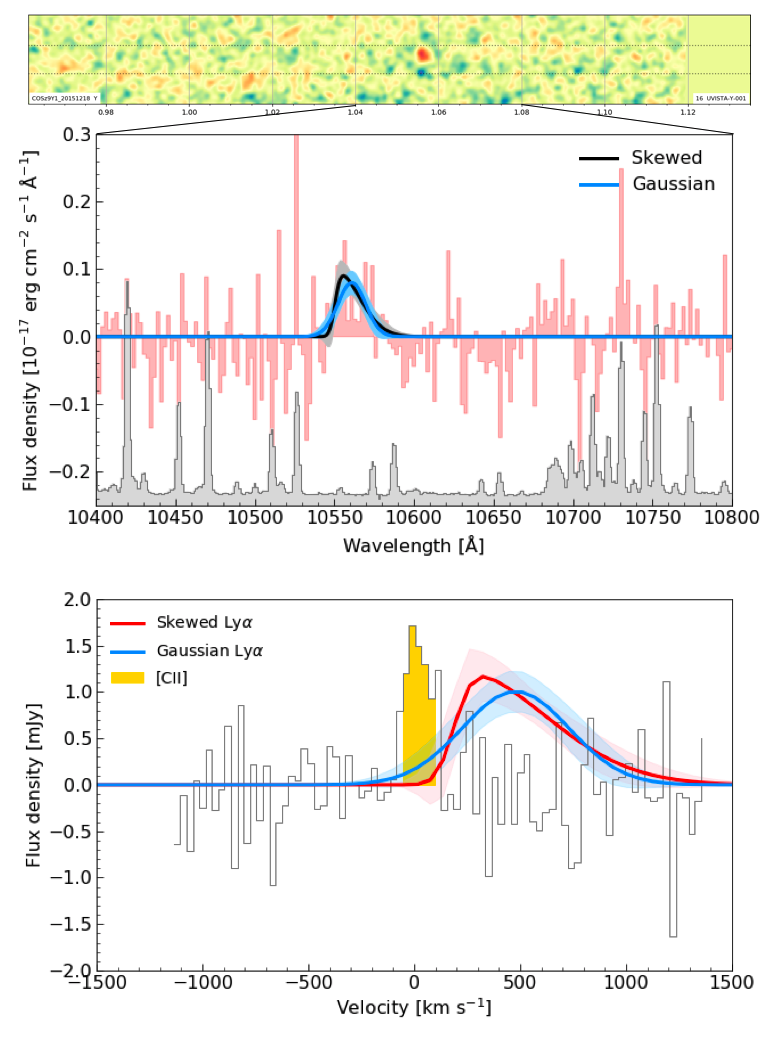}
    \caption{
    \textit{Top:} Bidimensional spectrum of \gal\ smoothed with a Gaussian filter for display purposes. \textit{Center:} The red and gray areas show the optimally extracted spectrum around the region of \lya\ emission at $z=7.677$ and the associated noise, binned by a factor of $\times 2$ for clarity. The blue and black lines and shaded areas indicate the best simple and skewed Gaussian models and their confidence intervals.
    \textit{Bottom:} The ALMA Band 6 spectrum of \gal\ where the velocity zeropoint corresponds to the systemic redshift of $z=7.677$ set by the \cii\ line. The golden area marks the channels used to optimize the S/N for the extraction. The blue and red lines and shaded areas indicate the \lya\ Gaussian models and their confidence interval as labeled, arbitrarily rescaled in flux density. We adopted the optical radial velocity definition.}
    \label{fig:spec}
\end{figure}

Scarce \cii\ detections were initially reported for \lya\ emitters at $z\gtrsim5$ and interpreted as owing to the low metal content and high ionization rates expected in early galaxies \citep{vallini_2015, harikane_2018}. This would also be reflected in bright \oiii\ lines and large \oiii/\cii\ ratios, then indeed reported \citep{hashimoto_2019}. Recent systematic searches for \cii\ in large samples at $z\sim4\text{--}8$ revealed a wide variety of galaxy properties and copious \cii\ photons from \lya\ emitters \citep[e.g.,][]{capak_2015, lefevre_2020, bouwens_2021, endsley_2022}. This allowed for the first statistical assessments of the correlations underlying the \lya\ and \cii\ emission.
Galaxies with large \lya\ equivalent widths (\ewlya) are found to correlate with a \cii\ deficit compared with the UV emission --- and, thus, SFR --- at $z>6$ \citep{carniani_2018, harikane_2018, hashimoto_2019}. This is in contrast with observations at $z\sim4\text{--}6$ \citep{schaerer_2020} where the difference might be driven by diverse sample selections, a possible redshift-dependent effect, and the large scatter in the properties of individual emitters \citep{harikane_2020}. The situation is further complicated by the observed clumpy structures and spatial offsets among the stellar light and gaseous or dusty components and even within different phases on the ISM \citep{bowler_2017, carniani_2018}. Therefore, the addition of new sources with multiple line transitions at the highest possible redshifts is still of primary importance to shed light on the earliest phases of galaxy formation and their impact on the surrounding IGM.\\

Here we report the detection of \lya\ emission from archival data of a Lyman-Break Galaxy (LBG) at the extreme bright end of the luminosity function with confirmed \cii\ emission at $z=7.677$ in the COSMOS field, dubbed \gal\ hereafter. This object has been indicated as a high-redshift photometric candidate (UVISTA-Y2 with $z_{\rm phot}=8.21^{+0.49}_{-0.50}$, \citealt{stefanon_2017, stefanon_2019}), later followed up with the Atacama Large Millimeter/submillimeter Array (ALMA) in a pilot \citep{schrouws_2021} and then in the full Reionization Era Bright Emission Line Survey (REBELS-36 with $z_{\rm phot}=7.88^{+0.58}_{-0.20}$, \citealt{bouwens_2021}). We describe our archival search and independent data reduction in Section~\ref{sec:data}. The analysis of the derived properties and a discussion of the relevant findings in the context of the current research landscape are presented in Section~\ref{sec:results}, followed by conclusions in Section \ref{sec:conclusion}. We assume a $\Lambda$-CDM cosmology with $\Omega_{\mathrm{m}} = 0.3$, $\Omega_{\Lambda} = 0.7$, and $H_0 = 70$ $\rm km\,s^{-1}\,Mpc^{-1}$. All magnitudes are expressed in the AB system.

\section{Data and methods}
\label{sec:data}

\subsection{Keck Near-infrared Spectroscopy}
\label{sec:keck}
As part of a large archival effort, we reduced spectroscopic $Y$-band data for \gal\ taken with the Keck Multi-Object Spectrometer For Infra-Red Exploration \citep[MOSFIRE,][]{mclean_2012}\footnote{Program ID: U043M, PI: G. Illingworth.}. The target was observed for 3.5 hours on December 18$^{\rm th}$ 2015, with an average seeing of ${\rm FWHM} = 0\farcs89$. We applied the distributed pipeline for the data reduction followed by an optimal extraction of the 1D spectrum \citep{horne_1986}. We corrected for the minor telescope drifting and derived the flux calibration based on a star observed within the same mask and in the same conditions as \gal. Finally, we anchored the photometry to the latest UltraVISTA DR4 $Y$-band flux as part of the COSMOS2020 catalog \citep{weaver_2022} to correct for further flux losses. We computed a reduced $\chi^2=1.0$ in regions of pure background in the 2D frame, confirming the reliability of the noise estimate. The 2D and 1D spectra are shown in Figure \ref{fig:spec}. 
We then modeled the \lya\ line profile as a Gaussian curve in the spectrum at the original resolution ($\sim31$ \kms\ velocity bin at the redshifted \lya\ wavelength). The best-fit parameters are reported in Table \ref{tab:data}. We also attempted to use a skewed Gaussian curve to allow for a possible asymmetry in the \lya\ profile (Figure~\ref{fig:spec}), but without significantly improving the fit (the AIC criterion does not strongly favor either model: $\Delta \rm AIC = 1.25$, with a $\sim50$\% probability that the skewed Gaussian is a better representation of the data). 
\begin{figure}
    \vspace*{0.3cm}
    \centering
    \includegraphics[width=\columnwidth]{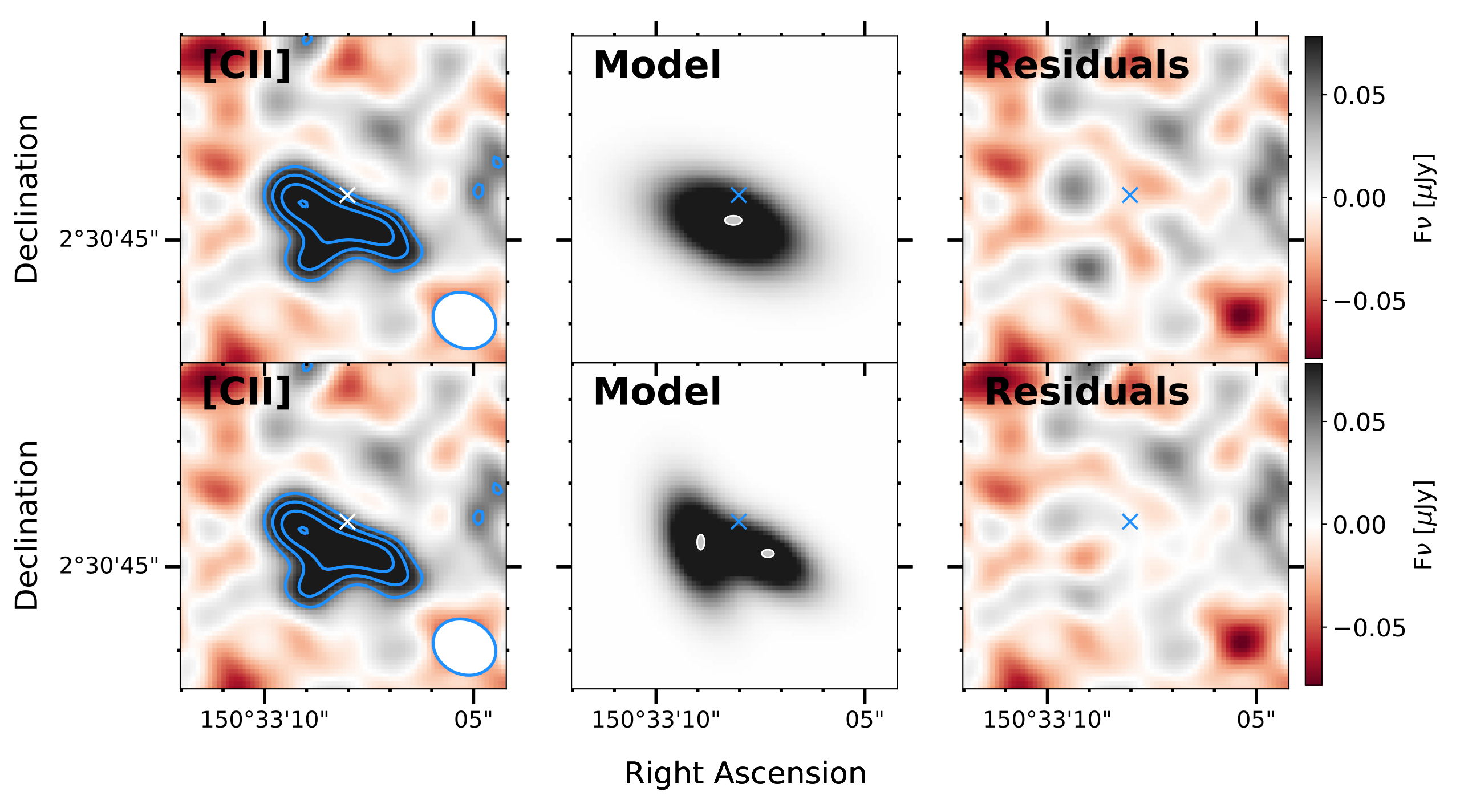}
    \caption{Modeling of the \cii\ emission with single (\textit{top}) and double (\textit{bottom}) elliptical Gaussians. The cross shows the location of the rest-frame UV and optical emission. The white ellipses show the locations of the peaks of the modeled \cii\ emission and their uncertainties. Blue lines in the left panels show 2--5$\sigma$ contours. The images are color scaled within $\pm3\times$rms.}
    \label{fig:imfit}
\end{figure}
\subsection{ALMA Far-infrared Spectroscopy}
\label{sec:alma}
\gal\ was observed in Band 6 with ALMA as part of the REBELS survey and its pilots\footnote{Program IDs: 2018.1.00236.S (PI: M. Stefanon), 2019.1.01634.L (PI: R. Bouwens).} (REBELS, \citealt{bouwens_2021, schrouws_2021}). 
The archival calibrated measurement set is provided by ESO and we used the Common Astronomy Software Applications package 
(\textsc{CASA}; \citealt{mcmullin_2007}) to analyze the data.  
We combined the available data with {\sc concat} and  produced a line cube and a continuum map with a pixel scale of $0\farcs1$ with {\sc tclean}, adopting a natural weighting scheme. The final beam size is $1\farcs59\times1\farcs29$, while we set a $30$ \kms\ channel velocity width for the cube, sufficient to resolve typical emission lines at high redshift over several bins.\\

In the bottom panel of Figure \ref{fig:spec}, we show the ALMA Band 6 spectrum extracted in an aperture optimizing its signal-to-noise ratio. We identified a significant line emission consistent with being \cii\ based on the identification of the \lya\ transition (Section \ref{sec:keck}). 
We measure a spectroscopic redshift of $z_{\rm sys,\,[CII]}=7.6771\pm0.0004$ from the Gaussian modeling of the detected line peaking at 219.027 GHz, in agreement the photometric redshift estimates (Table \ref{tab:data})\footnote{In parallel with our discovery of this line, an independent search by \citet{bouwens_2021} also find a \cii\ line for 
this source at $7.677$ (with a nominal S/N of $\sim7.8$).}.
We then collapsed the line detected channels and produced a \cii\ velocity-integrated line map (Figure~\ref{fig:imfit}), where the peak pixel emission (per beam) is detected at $\sim5.3\text{--}\sigma$ significance. We also found that the \cii\ line is spatially extended. We thus modeled the spatial profile of the \cii\ emission in the velocity-integrated map with {\sc imfit} as a single elliptical Gaussian, recovering the total flux over the channels maximizing the S/N (Figure \ref{fig:imfit}). The source is resolved and extended over $2\farcs43\times0\farcs93$ (deconvolved by the beam, Table~\ref{tab:data}). However, the \cii\ map shows two separated and similarly bright peaks, which might suggest the possible existence of sub-structures. We thus attempted a double-Gaussian fit that returned an identical total \cii\ flux estimate from two point-like sources separated by $\sim1\farcs6$. Higher spatial resolutions and S/N will be necessary to confirm and investigate possible clumping.\\

We did not detect any dust continuum emission from \gal, as also previously reported by \citet{schrouws_2021} based on slightly shallower observations from a pilot program for the REBELS survey. We, thus, set a $3\text{--}\sigma$ upper limit of 27 $\mu$Jy/beam at (observed) $1.44$ mm. Given the large beam size, this is computed as the pixel rms in the continuum map in units of Jy/beam obtained by combining all the existing observations at these wavelengths.

\subsection{Photometry}
\label{sec:photometry}
\begin{figure*}
    \vspace*{0.3cm}
    \centering
    \includegraphics[width=\textwidth]{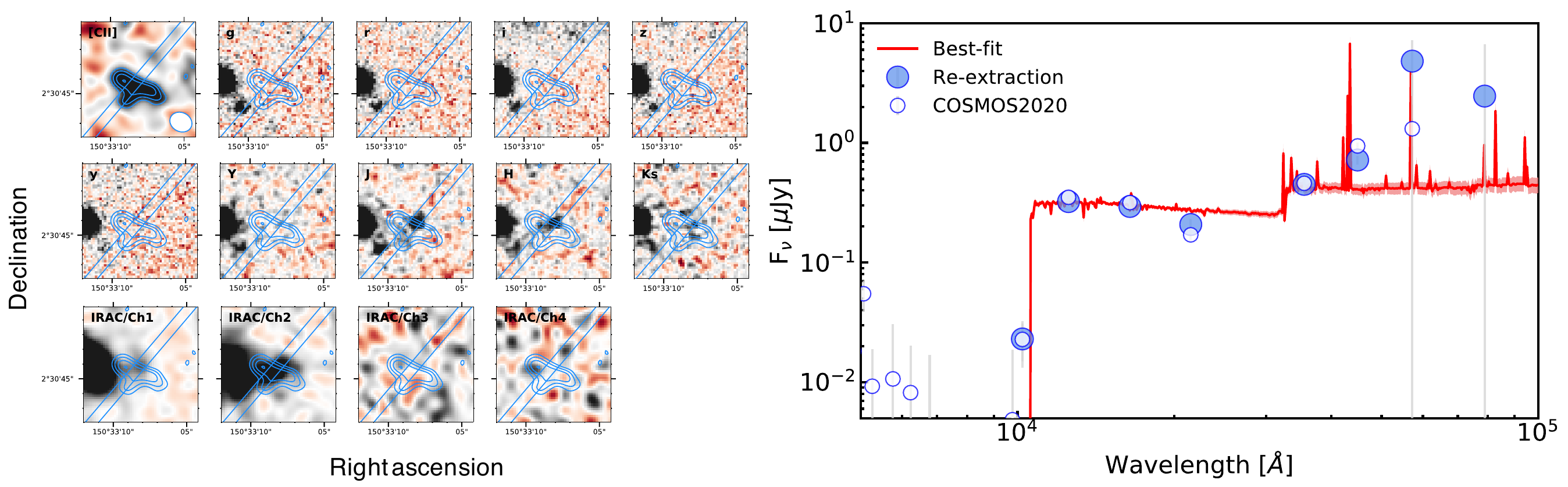}
    \caption{
    \textit{Left:} $8"\times8"$ cutouts of ALMA/\cii\ and optical to near-IR images from the COSMOS2020 catalog \citep{weaver_2022}. The cutouts are color-scaled within $\pm3\times$rms in each image and are  aligned N-E. The blue contours and slanted lines show the ALMA/\cii\ $2\text{--}5\sigma$ levels and the Keck/MOSFIRE slit orientation. The size of the ALMA beam is shown in the first cutout. 
    \textit{Right:} Best-fit photometric model of the SED with {\sc Bagpipes} at fixed $z=7.677$ and its uncertainty (red line and shaded area). Empty and filled blue circles mark the photometry from the COSMOS2020 catalog \citep{weaver_2022} and our custom re-extraction.}
    \label{fig:sed}
\end{figure*}

We extracted the photometry in the $Y, J, H, K_{\mathrm{s}}$ bands from the UltraVISTA DR4 and the four \textit{Spitzer}/IRAC channels from the latest reprocessed images \citep{weaver_2022} with the custom code {\sc Golfir}\footnote{\url{https://github.com/gbrammer/golfir}} (Kokorev, Brammer in preparation). The code uses priors from the highest spatial resolution imaging available to iteratively model and deblend the IRAC photometry. In Figure \ref{fig:sed}, we show the cutouts in optical to near-IR bands and a comparison between our photometry and that of the COSMOS2020 catalog extracted with {\sc The Farmer} (Weaver et al. in preparation). Generally, the two photometric sets are in good agreement, with the exception of the more conservative uncertainties on the IRAC fluxes that we adopt here.\\

We modeled our custom photometry with {\sc Bagpipes} \citep{carnall_2018} following the approach described in \citet{strait_2021} (Figure \ref{fig:sed}). We adopted the Binary Population and Spectral Synthesis templates \citep[BPASS,][]{eldridge_2009} and included nebular continuum and emission lines computed with {\sc Cloudy} \citep{ferland_2017}, leaving the ionization parameter free to vary in a range of $\mathrm{log}(U)=[-4, -1]$. We used a broken power law initial mass function with $\alpha=-2.35$ between $0.5\text{--}300$ \msun\ and $\alpha=-1.3$ at $0.1\text{--}0.5$ \msun\footnote{The fiducial BPASS IMF has a similar shape to that of \citet{chabrier_2003} or \citet{kroupa_2001}. The resulting mass-to-light ratio tracks well the value for \citet{bruzual_2003} single population synthesis models for ages of $<6$ Gyr \citep[Figure 7 in][]{stanway_2018}.}. We used a flexible exponential star formation history able to rise, decline, or staying constant, along with a \citet{calzetti_2000} dust attenuation law ($A_{\rm V}=0\text{--}3$) where twice as much dust is assigned to \hii\ regions as in the diffuse ISM in the first 10 Myr. We left the metallicity as a free parameter ($Z=0.005\text{--}5\,Z_{\odot}$), but fixed the redshift to its spectroscopic value $z_{\rm sys, [CII]}$. A detailed description of the modeling, its biases, and systematic uncertainties can be found in \citet{strait_2021}.\\

The best-fit parameters are reported in Table \ref{tab:data}. From the best-fit model, we also derived the continuum emission underlying the \lya\ line since no detectable trace was found in the MOSFIRE spectrum. This estimate is consistent with that derived from the observed continuum in $J$ band (the closest filter completely redward of the \lya\ break) and by assuming a flat constant spectrum in $F_{\rm \nu}$.
We further computed the unobscured SFR from the rest-frame UV luminosity at 1500 \AA\ as $\mathrm{SFR_{UV}} [M_\odot\,\mathrm{yr}^{-1}] = 8.24\times10^{-29}\,L_{\rm \nu, UV}\, \mathrm{[erg\,s^{-1}\,Hz^{-1}]}$ \citep{schaerer_2020}. Moreover, we derived an upper limit on the obscured SFR from the total infrared luminosity $L_{\rm IR(8\text{--}1000\,\mu m)}$ obtained by rescaling a modified black body curve with a dust temperature of $T_{\rm dust}=45\,\mathrm{K}$ and a $\beta_{\rm IR}=1.5$ power-law exponent to the continuum emission at $\sim160\,\mu$m rest-frame, modeling the effect of the CMB as in \citet{dacunha_2013}. We then applied the conversion $\mathrm{SFR_{IR}}\,[M_\odot\,\mathrm{yr}^{-1}] = 1.4\times10^{-10} L_{\rm IR}\,[L_\odot]$. For a dust mass absorption coefficient of $k_0=8.94$ g cm$^{-2}$ at $158$ $\mu$m \citep{hirashita_2014}, the $3\text{--}\sigma$ upper limit on the continuum detection corresponds to a dust mass of $M_{\rm dust}<10^{6.9}\,M_\odot$. The ratio of \lcii\ and its underlying dust continuum luminosities is consistent with the upper end of the distribution of the 158\,$\mu$m continuum detected sources in the REBELS sample \citep{sommovigo_2022}.

\begin{deluxetable}{lr}
  \tabletypesize{\normalsize}
  \tablecolumns{2}
  \tablecaption{Physical properties of \gal.\label{tab:data}}
  \smallskip
  \tablehead{}
   \startdata
    %
    R.A. (J2000) & 10:02:12.54 (150.55224056 deg) \\
    Decl. (J2000) & +2:30:45.9 (2.51273892 deg) \\
    \rule{0pt}{4ex}$z_{\rm phot,\,C20}$\tablenotemark{a}& $8.31^{+0.31}_{-0.58}$\\
    %
    $z_{\rm sys, [CII]}$& $7.6771 \pm 0.0004$\\  
    %
    $M_{\rm UV,\,1500}$&  $-21.95\pm0.04$\\
    $\beta$& $-2.58\pm0.16$\\
    %
    $\mathrm{log}(M_\star [M_\odot])$ & $9.67^{+0.09}_{-0.15}$\\
    %
    $A_{\rm V}$ [mag] & $0.18^{+0.05}_{-0.05}$ \\ 
    $Z$ [$Z_\odot$] & $0.29^{+0.18}_{-0.10}$ \\ 
    %
    SFR(UV) [$M_\odot\,\mathrm{yr}^{-1}$]& $22\pm1$\\
    SFR(IR) [$M_\odot\,\mathrm{yr}^{-1}$]& $<15$\\
    %
    \rule{0pt}{4ex}$L_{\rm Ly\alpha}$ [$\mathrm{erg\,s^{-1}}$]& $(1.30\pm0.28)\times10^{43}$ \\ 
    \ewlya\ [\AA]& $24^{+5}_{-6}$ \\ 
    $\sigma_{\rm vel}$(\lya) [$\mathrm{km\,s^{-1}}$]& $248\pm18$ \\ 
    %
    \dlya\ [$\mathrm{km\,s^{-1}}$]& $480\pm67$ \\ 
    \dxya\ [arcsec]& $0.64^{+0.11}_{-0.12}$ \\ 
    %
    \rule{0pt}{4ex}$S_{\rm [CII]}$ [\jykms]& $(0.33\pm0.07)$ \\ 
    $L_{\rm [CII]}$ [$L_\odot$]& $(4.41\pm0.90)\times10^{8}$ \\ 
    $L'_{\rm [CII]}$ [\kkmspc]& $(2.00\pm0.41)\times10^{9}$ \\ 
    $\sigma_{\rm vel}$(\cii) [$\mathrm{km\,s^{-1}}$]& $63\pm13$\\
    $\rm FWHM_{maj}$(\cii) [arcsec]& $2.43\pm0.63$\\
    $\rm FWHM_{min}$(\cii) [arcsec]& $0.93\pm0.44$\\
    %
    $S_{\rm 1.44mm}$ [$\mu$Jy]& $<27$ \\ 
    $\mathrm{log}(L_{\rm IR,\,45K}/L_\odot)$& $<11.03$ \\ 
     \enddata
  \tablecomments{Upper limits at $3\sigma$.}
  \tablenotetext{a}{Photometric redshift from \textsc{Farmer+LePhare} in \citet{weaver_2022}. An updated run with {\sc Eazy} adopting more conservative uncertainties on IRAC fluxes as the one used in this work returns $z_{\rm phot,\,Cl} =7.77^{+0.35}_{-0.06}$. The estimate based on the ``classical'' aperture photometry is $z_{\rm phot,\,Cl} =7.69^{+1.06}_{-0.15}$.}
\end{deluxetable}

\section{Results and discussion}
\label{sec:results}

\subsection{\lya\ and \cii\ offsets}
\label{sec:offset}
In Figure \ref{fig:spec}, the velocity offset of the \lya\ line with respect to the systemic redshift set by the \cii\ emission can be fully appreciated. The peak of the Gaussian profile is redshifted by \dlya$\sim500$ \kms\ from $z_{\rm sys,[CII]}$ with a significant high-velocity tail extended up to $\sim1000$ \kms\footnote{For the sake of completeness, we retrieve $\Delta v_{\mathrm{Ly\alpha}}=(182\pm57)$ \kms\ from the skewed Gaussian model. The (non-parametric) first moment computed over the wavelength bins where $F_{\rm \lambda}>1$\% of the peak derived from the best-fit Gaussian model ($\lambda \in [10535.3, 10587.4]$ \AA) and weighted by the S/N to reduce the impact of noise peaks is $432$ \kms.}. This value is $\sim3.5\times$ higher than the observed average offset for similarly bright objects at $z\sim6\text{--}7$ ($\sim150\text{--}200$ \kms, \citealt{pentericci_2016,matthee_2019}) and the recent modeling of the \muv--\dlya\ relation at $z=7\text{--}8$ by \citealt{mason_2018_igm_neutral}, computed without including outflow models explicitly (Figure \ref{fig:correlations}). However, \dlya\ is consistent with the average estimate for $M_{\rm UV}<-22$ galaxies at $z>6$ reported by \citet{endsley_2022}, including 4 objects from the {\sc Rebels} survey\footnote{The average estimate from \citet{endsley_2022} assumes \dlya$=504\pm52$ \kms\ for WMH5(-B) at $z=6.07$, in lieu of the $265\pm52$ \kms\ value for WMH5(-A) listed by \citet{matthee_2019} and adopted here \citep{willott_2015}.}, similar to CLM1 at $z=6.17$ \citep{willott_2015}, but lower than what reported for B14-65666 at $z=7.15$ (\dlya$\sim770$ \kms, \citealt{hashimoto_2019}).
Similarly to CLM1, but at odds with B14-65666, in the case of \gal\ the velocity offset does not pair with a remarkably low value of the \ewlya\ and high \lcii, despite the similar \muv\ (Table \ref{tab:data}). Quite the opposite: \gal\ is among the strongest UV-bright \lya\ emitters at $z>6$ with an \ewlya$=24^{+5}_{-6}$ \AA\ (Figure \ref{fig:correlations}). The velocity offset is consistent with the average for LBGs at similar magnitudes at $z\sim2\text{--}3.5$ \citep{erb_2004, steidel_2010, willott_2015, marchi_2019, cassata_2020}.\\ 

We also report a spatial offset of $\Delta x_{\rm Ly\alpha} = 0\farcs64^{+0.11}_{-0.12}$ ($3.16^{+0.54}_{-0.59}$ kpc) between the peaks of the rest-frame UV stellar light --- co-located with the \lya\ emission --- and the one of the \cii\ emission. Larger spatial offsets ($\sim0\farcs75\text{--}1\arcsec$) would be measured from the peaks of the possible sub-structures in the \cii\ map, if confirmed (Figure \ref{fig:imfit}). At the current spatial resolution at rest-frame UV and FIR wavelengths, an overlap between the stellar and cold gaseous components remains.

\subsection{A hint of complex geometry}
\label{sec:geometry}
The spatial offset and the large \lya\ velocity shift could be explained if we considered \gal\ as composed by multiple merging clumps with different physical properties \citep{carniani_2018, hashimoto_2019}. The lack of \lya\ photons percolating from the region close to the peak of the \cii\ emission could be explained by a large amount of neutral hydrogen resonantly trapping the \lya\ light and a minimal mass of dust absorbing it. Following \citet{heintz_2021} and assuming that \cii\ traces the \hi\ mass in the ISM at $z>6$, we derived a total $\mathrm{log}(M_{\rm H\scriptscriptstyle{I}}/M_\odot)_{Z=0.3Z_\odot}=(10.58 \pm0.17)$ for a representative metallicity of $Z=0.3\,Z_\odot$ and including the uncertainties on the calibration. The \hi\ mass is $\sim8\times$ higher than \mstar\ in \gal\ as a global system, suggesting the existence of large neutral gas reservoirs. Under these assumptions, we could consider this ratio as lower limit consistent with the extrapolation of the \mhi/\mstar\-$z$ trend derived up to $z\sim6$ \citep[2022 in preparation]{heintz_2021}. We further derived a column density of $N_{\rm H\scriptscriptstyle{I}}\sim10^{22.7\pm0.3}$ cm$^{-2}$ within an effective ellipse of $43.4\pm23.8$ kpc$^2$ enclosing half of the total \cii\ luminosity as estimated from the Gaussian modeling (half-light semi-axes $a_{\rm maj,\,min} = \mathrm{FWHM_{maj,\,min}/2}$). This \nhi\ estimate is consistent with the values in the high-density tail from direct observations of gamma-ray bursts in star-forming galaxy cores up to $z\sim4\text{--}5$ \citep{tanvir_2019}. These column densities are challenging for the predictions of a simple expanding shell model \citep{verhamme_2015}. In this case, the \lya\ emission would not be able to escape from the \cii-emitting dense \hi\ region even with the low amount of dust expected \gal\ from the upper limit on the continuum detection (Section \ref{sec:photometry}). If some of the observed \lya\ light were resonantly scattered from the \cii\ clump and escaping at high velocities, we would expect a simultaneous decrease in \ewlya, which is not observed. Moreover, the overlap between the \lya\ and UV emission lends support to an in-situ production scenario. Different levels of metal enrichment, hardness of the ambient radiation field, and dust extinction could then concur to explain the spatial offsets \citep{carniani_2018}.\\

This is a simple calculation providing an average and approximate estimate of the \hi\ column density from \cii, critically depending on the metallicity of the ISM. Here we adopted the loose constraint from the SED modeling of the UV-bright component, but indirect arguments support this choice. On the one hand, discounting very high dust temperatures, the non-detection of continuum emission suggests $Z<Z_\odot$, as one could expect at such high redshifts (but see \citealt{watson_2015}). On the other hand, lower metallicities and extreme \mhi\ are disfavored by the approximate estimate of the dynamical mass $M_{\rm dyn} \approx 1.16\times10^{5} v_{\rm circ}^2 D \sim 3\times10^{10} M_\odot$ that we computed from the \cii\ emission \citep{wang_2013}. Here $v_{\rm circ}=0.75\,\mathrm{FWHM_{\rm [CII]}} / \mathrm{sin}(i)$ is the maximum circular velocity of the gas in \kms\ assuming a rotating disk geometry for the \cii-emitting component, $D=1.5\times\mathrm{FWHM_{\rm maj}}$ the size in kpc, and $i=\mathrm{cos}^{-1}(a_{\rm min}/a_{\rm maj})$ is the disk inclination from the minor/major axis ratio.\\ 

\subsection{A bright system deep in the reionization epoch}
\begin{figure*}
    \vspace*{0.3cm}
    \centering
    \includegraphics[width=\textwidth]{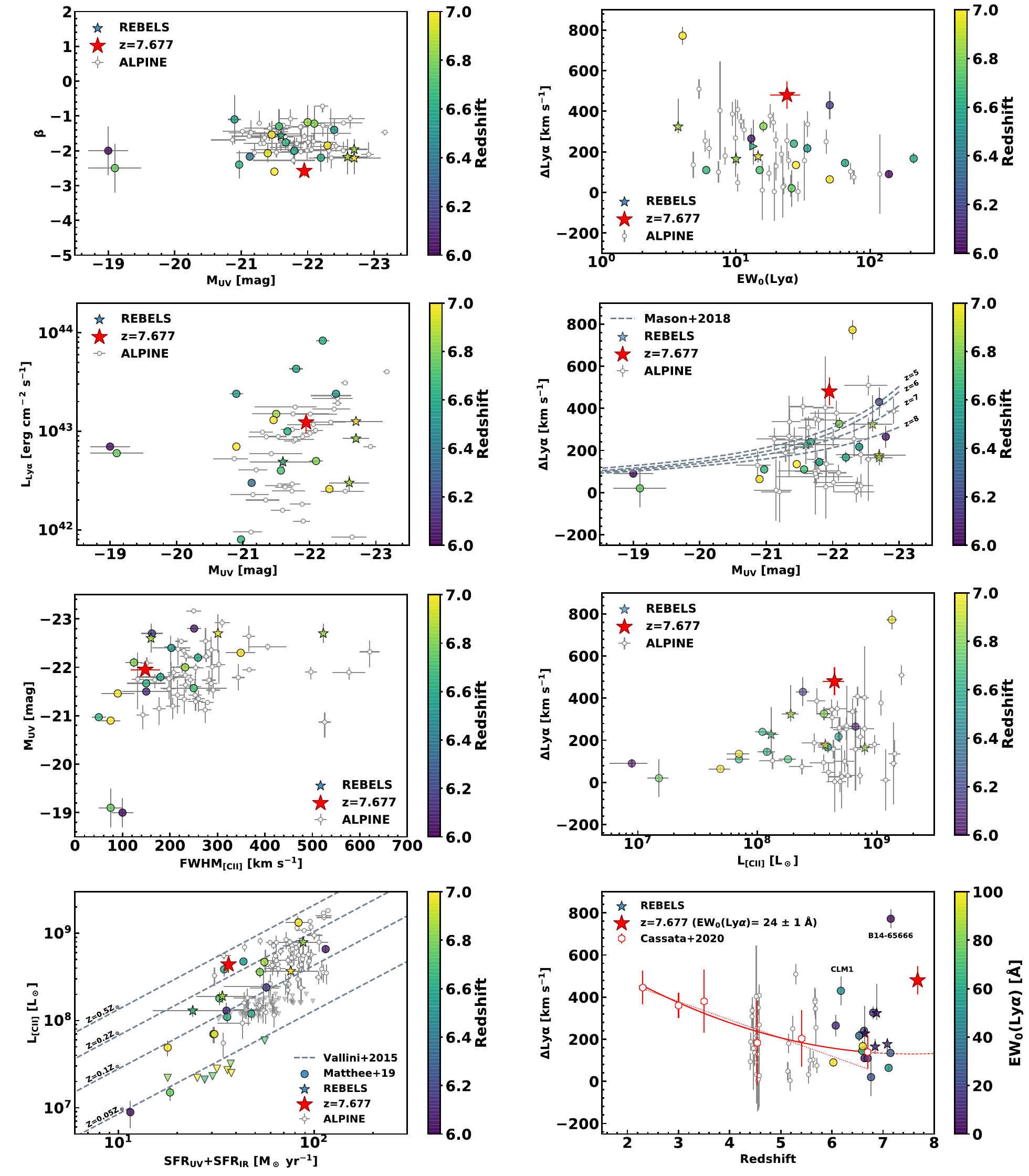}
    \caption{The location of \gal\ in the plane of observables. 
    \textit{Left column:} 
    $M_{\rm UV}$ vs. $\beta$;
    $M_{\rm UV}$ vs. $L_{\rm Ly\alpha}$;
    $\mathrm{FWHM_{\rm [CII]}}$ vs. $M_{\rm UV}$;
    $\rm SFR_{UV}+SFR_{IR}$ vs. $L_{\rm [CII]}$.
    \textit{Right column:} 
    $\mathrm{EW_0(Ly\alpha)}$ vs. \dlya;
    $M_{\rm UV}$ vs. \dlya;
    $L_{\rm [CII]}$ vs. \dlya;
    redshift vs. \dlya. 
    In each panel, the red filled star marks the location of \gal. The filled circles show the compilation of $z\sim6\text{--}7$ \lya\ and \cii\ emitters from \citet{matthee_2019} and \citet{hashimoto_2019} color-coded as labeled. Filled stars indicate the galaxies from the REBELS survey with \lya\ detection presented in \citet{endsley_2022} ($\beta$ values from \citealt{bouwens_2021}). Sources from the ALPINE survey at $z\sim4\text{--}6$ are shown in gray \citep{cassata_2020,faisst_2020,bethermin_2020}.
    The dashed lines in the bottom-left panel mark the SFR-\lcii\ relation as a function of metallicity as in \cite{vallini_2015}.  The ones in the \muv-\dlya\ panel show the median relation as a function of redshift in \cite{mason_2018}. The red symbols and lines in the bottom-right panel indicate the average values and the best-fit linear and parabolic trends as in \cite{cassata_2020}.}
    \label{fig:correlations}
\end{figure*}

The best-fit SED modeling at fixed $z_{\rm sys,\,[CII]}$ suggests that \gal\ is a relatively massive ($\mathrm{log}(M_\star/M_\odot)=9.67^{+0.09}_{-0.15}$) and UV-bright system ($2\times$ brighter than 
$M^{\star}_{\rm UV}$ at $z=7\text{--}8$, \citealt{bouwens_2021_LF}). 
For the sake of comparison with the literature and considering \gal\ as a single system, the SFR(UV)+SFR(IR) and \lcii\ estimates place it on the upper end empirical relation among these quantities observed across redshifts (Figure \ref{fig:correlations}, \citealt{delooze_2014, vallini_2015, schaerer_2020}). Also, \gal\ does not deviate significantly from the correlations observed among \muv, $\beta$-slope, FWHM(\cii), and $L_{\rm Ly\alpha}$ (Figure \ref{fig:correlations}).\\

The large velocity offset between the \lya\ peak and the systemic redshift is what distinguishes \gal\ compared with most of the known high-redshift galaxies, in a similar fashion as for the systems reported by \citet{willott_2015} and \citet{hashimoto_2019}. Empirical correlations among \dlya\ and \ewlya, \lcii, and \muv\ appear to be in place at $z>6$ based on composite literature compilations \citep{hashimoto_2019, endsley_2022}. Since \lya\ is a resonant line, its velocity offset can be interpreted as owing to large \hi\ column densities (paired with low \ewlya) and as a signature of outflowing material \citep{verhamme_2006, verhamme_2015}. The empirical correlations among \dlya\ and \lcii\ and \muv\ could just reflect larger \nhi\ and stronger SFR-driven outflows in bright \cii\ and UV emitters, respectively \citep{hashimoto_2019}. Discounting luminosity selection effects, the absence of sources detected at high \dlya\ for low \lcii\ and faint \muv\ would be explained by the fact that \lya\ photons can escape closer to their rest-frame wavelength. However, in Figure \ref{fig:correlations} we show that the addition of the ALPINE sample of main-sequence galaxies at $z\sim4\text{--}6$ does not fully support these findings. While this could be due to the different selection functions \citep{harikane_2020}, a transition around $z\sim6$ could be expected due to the increasing IGM neutral fraction at $z>6$. In addition, extremely large \dlya\ could simply reflect the existence of merging substructures or galaxies, calling in question the physical association of \cii, UV, or \lya\ emitting clumps -- for which correlations might not a priori exist.\\

\subsection{A large reionization patch?}
\label{sec:bubble}
The existence of an ionized bubble around \gal\ could facilitate the escape of \lya\ photons. The detection of strong \lya\ emission (\ewlya$=24^{+5}_{-6}$ \AA) might, in fact, suggest the existence of such a large patch of reionization \citep{stark_2017, leonova_2021}. The measured \ewlya\ is $\times 2$ larger than the median values for galaxies with $M_{\rm UV}<-20.4$ in the field at $z\sim7$, but consistent with the value reported for an overdensity of emitters at a similar redshift \citep{endsley_2021b}. However, a model of the observed \ewlya\ distribution in an inhomogeneously reionizing IGM by \cite{mason_2018} (their Figure 7) returns a probability of $5.8^{+1.0}_{-1.1}$\% to detect a galaxy with $M_{\rm UV}=-22$ and such large \ewlya\ at $z=7.6$, given the current constraints on the IGM neutral fraction at that redshift ($x_{\rm H\,\scriptscriptstyle{I}}=0.74^{+0.07}_{-0.06}$, \citealt{mason_2019b, hoag_2019}). This points once more to the prominent role of large \lya\ velocity offset in facilitating the photon escape. If future observations confirmed \dlya\ of the order of $\sim500$ \kms\ (Table \ref{tab:data}), it would be potentially sufficient to explain the emission without invoking the transparency of the surrounding IGM over large scales \citep{endsley_2022}.

\section{Conclusions}
\label{sec:conclusion}
We reported the archival discovery of strong \lya\ emission from a UV-bright ($2 \times$ brighter than $M^{\star}_{\rm UV})$ system spectroscopically confirmed by its \cii\ emission detected with ALMA at $z=7.677$. The peaks of the \lya\ and \cii\ emission appear offset both spatially ($\sim3$ kpc) and in velocity (\dlya\ $\sim500$ \kms), with the emission from the recombining hydrogen atoms being coincident with the rest-frame UV stellar light.\\ 

The velocity offset is among the largest reported so far and $3.5\times$ higher than the average at $z\sim6\text{--}7$, while being paired with a large \ewlya\ $=24^{+5}_{-6}$\AA. The spatial and velocity offsets hint at the existence of a complex structure and merging sub-units. The distribution of large amounts of \hi\ gas, coupled with a minimal amount of dust (as derived from the $3\text{--}\sigma$ upper limit on the continuum emission at $160$ $\mu$m rest-frame, $M_{\rm dust}<10^{6.9}$ \msun), could help explaining the observed spatial offsets. By assuming that \cii\ primarily traces \hi, we estimated $\mathrm{log}(M_{\rm HI}/M_\odot)_{Z=0.3Z_\odot}\sim 10.6$, $\sim8\times$ larger than \mstar\ estimated for \gal\ as a single system. For a simplified uniform distribution of the measured size of the system, we derive an approximate estimate of \nhi\ in excess of $10^{22.5}$ cm$^{-2}$, enough to trap and absorb the \lya\ emission from the \cii-emitting clump. Escape at very high velocity (possibly suggested by the observed \dlya$\sim500$ \kms) would be hardly compatible with the large \ewlya\ that we estimate.\\

As a single system, \gal\ stands out compared with other high-redshift systems for the large \lya\ velocity offset and \ewlya. The latter might be a common feature among bright UV-galaxies at $z\sim6\text{--}7$, as supported by the observed correlations between \dlya, \lcii\ ($\propto$\mhi), and \muv\ ($\propto$ SFR$_{\rm UV}$). However, such correlations are not immediately evident when including samples at $z\sim4\text{--}5$. Moreover, extremely large \dlya\ might result from spatially disconnected regions, challenging the strength of the observed correlations in absence of high-resolution observations. Finally, such a large \ewlya\ from a system as bright as $M_{\rm UV}=-22$ has a relatively low probability of $\sim5$\% of being observed at $z=7.6$, given the current constraints on the IGM neutral fraction. This independently suggests the central role of outflows in facilitating the escape of \lya\ photons, besides the possible existence of a ionized bubble around this system.\\ 

Several aspects deserve to be clarified in future work. High-resolution and deeper ALMA \cii\ observations are necessary to unambiguously confirm and describe the
clumpy substructure of \gal, currently only suggested by our modeling of the \cii\ map. In addition, the emission of several rest-frame optical lines from the ionized gas phase (oxygen, hydrogen) is within reach of the \textit{James Webb Space Telescope}. High-resolution measurements could unambiguously determine the spatial distribution, velocity offset, \lya\ escape fractions, metallicities, and ionization parameter of the ionized phases and the stellar component.      

\section{Acknowledgements}
\label{ref:acknowledgements}
We acknowledge the constructive comments from the anonymous referee that improved the content and presentation of the results. 
We warmly thank Paolo Cassata for providing data from his work, Rychard Bouwens for his suggestions about the REBELS survey, and Ivo Labb\'{e} for his useful comments on this analysis.
The Cosmic Dawn Center (DAWN) is funded by the Danish National
Research Foundation under grant No.~140. F.V. and K.E.H. acknowledge
support from the Carlsberg Foundation under grants CF18-0388 and
CF21-0103. K.E.H. acknowledges support by a Postdoctoral Fellowship
Grant (217690--051) from The Icelandic Research Fund. S.F. and
D.W. are supported by Independent Research Fund Denmark grant
DFF--7014-00017. C.M. acknowledges support by the VILLUM FONDEN under
grant 37459. S.T. and J.R.W. acknowledge support from the European
Research Council (ERC) Consolidator Grant funding scheme (project
ConTExt, grant No.~648179). Some of the data presented herein were
obtained at the W. M. Keck Observatory, which is operated as a
scientific partnership among the California Institute of Technology,
the University of California and the National Aeronautics and Space
Administration. The Observatory was made possible by the generous
financial support of the W. M. Keck Foundation. The authors wish to
recognize and acknowledge the very significant cultural role and
reverence that the summit of Maunakea has always had within the
indigenous Hawaiian community.  We are most fortunate to have the
opportunity to conduct observations from this mountain. This paper
makes use of the following ALMA data: 2018.1.00236.S,
2019.1.01634.L. ALMA is a partnership of ESO (representing its member
states), NSF (USA) and NINS (Japan), together with NRC (Canada), MOST
and ASIAA (Taiwan), and KASI (Republic of Korea), in cooperation with
the Republic of Chile. The Joint ALMA Observatory is operated by ESO,
AUI/NRAO and NAOJ.  
\bibliography{bib_highz}
\bibliographystyle{aasjournal}

\end{document}